\documentclass[11pt]{article}

\def\ben{\begin{enumerate}} \def\een{\end{enumerate}}
\def\beq{\begin{equation}} \def\eeq{\end{equation}}
\def\beqn{\begin{equation*}} \def\eeqn{\end{equation*}}
\def\bea{\begin{eqnarray}} \def\eea{\end{eqnarray}}
\def\ba{\begin{array}} \def\ea{\end{array}}
\def\beann{\begin{eqnarray*}} \def\eeann{\end{eqnarray*}}
\def\beasn{\begin{sneqnarray}} \def\eeasn{\end{sneqnarray}}

\begin{document}

\section*{Book Review}

\subsection*{{\bf The Genesis of General Relativity.} {\it J\"urgen Renn, ed.} Springer, Dordrecht. 2007}
\subsubsection*{Volume 1 {\bf Einstein's Zurich Notebook: Introduction and Source.} {\it By Michel Janssen, John Norton, J\"urgen Renn, and John Stachel}, 487 p.}
\subsubsection*{Volume 2 {\bf Einstein's Zurich Notebook: Commentary and Essays.} {\it By Michel Janssen, John Norton, J\"urgen Renn, Tilman Sauer, and John Stachel}, 451 p.}
\subsubsection*{Volume 3 {\bf Gravitation in the Twilight of Classical Physics: Between Mechanics, Field Theory, and Astronomy.} {\it J\"urgen Renn and Matthias Schemmel, eds.}, 619 p.}
\subsubsection*{Volume 4 {\bf Gravitation in the Twilight of Classical Physics: The Promise of Mathematics.} {\it J\"urgen Renn and Matthias Schemmel, eds.}, 533 p.}

These volumes are the result of over two decades of effort, by most of the leading scholars in the field, to understand the process that culminated in 1915 and 1916 in Einstein's publication of the general theory of relativity. In addition to relativity physicists 
the project involved, both individually and more frequently collaboratively, historians and philosophers of science. The central objective was, through this richly documented case study, to identify universal features of the epistemological transformation that the authors have called a ``Copernican process": How is it that heuristic guides can  render conceptual changes that invalidate their use? This dynamical transmutation is firmly rooted in received societal and disciplinary scientific knowledge. In the particular case under study here, most relativists will probably have little trouble rejecting the mistaken popular notion that Einstein was an isolated genius, creating his new world through shear inspired imagination. But I wager that most of us will be surprised to learn, as I have through reading these volumes, how steadfast was Einstein in applying the principles and methods that lay at the standard core of early twentieth century theoretical physics. He did not act alone. His heuristics differed from the rest of the theoretical physics community mainly in the priority he assigned to the equivalence principle. I would venture to say if we wish to identify a trait of genius, it would be the persistence with which he held to this fundamental insight.

There is an obvious utilitarian reason why this study should be of interest to us: the better we understand the process of scientific knowledge acquisition, the better we will be able to create conditions in which young scientists can follow the lead of innovators like Albert Einstein. But there are additional attractions having to do with the practice of our own craft. We are not that frequently cognizant of the conceptual matrix in which we work. The authors borrow the notion of ``mental models'' from the cognitive sciences. And we are normally even less able to escape this contemporary framework and view the world as it was modeled by our predecessors. That act has the potential of opening new vistas. In fact, to smugly dismiss the work of our predecessors as simply errant science would stultify our own practice. Yet, how many of us know of the work of, for example, Max Abraham, Gustav Mie, Gustav Herglotz, and Gunnar Nordstr\" om, just to name a few of the significant scientists who feature in these volumes? 

I will review here several of the key steps that Einstein undertook starting in 1907. The objective is twofold. I want to make the point that this process bears no resemblance to the manner in which general relativity is presented in modern textbooks. Furthermore the steps I have selected illustrate procedures that Einstein generalized in his search process. They thus provide technical and physical examples of the dual strategy that has been convincingly identified by the authors. He had been commissioned to write a review article addressing the larger implications of the special theory of relativity. 

The story begins with the well-known "happiest thought of my life", Einstein's realization that just as with the magnetic field under a change of inertial frame, the gravitational field can be made to vanish under the transformation to an accelerating frame of reference. Already in 1907 Einstein used the equivalence principle to deduce that clocks at higher gravitational potential (in a uniform gravitational field) run at a faster rate than same clocks at lower potential (when they are brought together and total elapsed time is compared). As is amply illustrated in these texts,  through original texts, correspondence, and historical analysis, soon other eminent physicists were attempting to reconcile gravitation with the special theory of relativity. Most notable among these players were Max Abraham  and Gustav Nordstr\"om.  Abraham devised a theory in which the speed of light was not constant, not recognizing initially that this could not be done in the context of the special theory. But the idea did apparently stimulate Einstein's thinking. Insisting that point particles undergo accelerations independent of their inertial masses, and that this four-acceleration corresponds to minus the gradient of a gravitational potential, a constant velocity of light leads to the unacceptable conclusion that the four-velocity must be perpendicular to the surfaces of constant potential, i.e., $\frac{d c^2}{d \tau} = \frac{d }{d \tau} \left( U_\mu U^\mu \right) = - 2 U^\mu \phi_{, \mu} = 0$. As John Norton illustrates in one of his contributions, if one then has a particle moving instantaneously horizontally with speed $v$  where there is a downward vertical field $-g$, the resulting vertical acceleration is $\frac{d v_z}{dt} = - g \left( 1 - v^2/c^2 \right)$.  For Einstein this was an unacceptable violation of the equivalence principle. Einstein concluded that there was no way of reconciling  the equality of inertial and gravitational mass with the fact that they must both transform as the zero component of a four vector under Lorentz transformations. He made his first step in generalizing the relativity principle, and thereby abandoning Lorentz covariance, in 1911 and 1912. He examined from the perspective of an observer undergoing constant acceleration relative to Minkowski space, the propagation of a pulse of light. Insisting that between any two fixed points in this accelerated frame the observed speed of light must be the same, no matter at what accelerated frame time the measurement is made, he deduced that the speed of light in the accelerated frame must depend on the location. In insisting on time independence he implemented an extended principle of relativity: the laws of physics must be covariant under uniformly accelerated changes of frame. Thus since among all of the solutions described by the new physical laws in the original unaccelerated frame there is known to exist a static uniform gravitational field with our familiar law of free fall, and therefore precisely the same solution with the same physical consequences must exist in the accelerated frame of reference. This was the sense he which he came to believe that the accelerated frame observer could legitimately maintain that she was ``at rest''. The result of this requirement was that the speed of light $c$ in a frame with spatially varying gravitational potential did indeed depend on the position. Furthermore, in examining the motion in the accelerated frame of a particle moving inertially in the Minkowski frame, he deduced that the equation of motion was $\frac{d}{dt} \left( \frac{\dot z}{c^2} \right)  = - \frac{1}{c} \frac{d \phi}{d z}$.  After then demonstrating that the energy of the particle of rest mass $m$ was equal to $m c$ he was able to deduce that the force that correctly augmented this energy was $\vec F = - m \vec \nabla c(z)$. Thus he began to interpret the variable speed of light as the gravitational potential.  In addition, continuing to work with the uniformly accelerated frame he noticed that if light moving in the horizontal direction diminishes its coordinate speed $\frac{dx}{dt}$ with increasing height $z$, yet ideal clocks (synchronized with ideal clocks at the origin of the Minkowski frame) must measure the constant speed $c$, then the proper elapsed time recorded by ideal clocks in the frame with potential $\phi = c(z)$ must register a duration $d\tau = c(z) dt$.  

I have gone through this reasoning in some detail so as understand Einstein's momentous next step. He had by this time adopted Minkowski's four dimensional geometric formulation of special relativity. Thus it was now natural for him to interpret the square of the variable speed of light as a spatially variable ``$0 0$'' component of a curved spacetime metric.  The challenge was then to appropriately formulate field equations for $c$. This was the situation when he arrived in Zurich in 1912 to take up his new position at his alma mater, the ETH. What transpires next constitutes the centerpiece of these four volumes. We are truly fortunate that a documented record of Einstein's ruminations exists for this transitional period leading to the general theory of relativity. Einstein's Zurich notebook has been known for decades, and it has even been published with limited annotation in the Collected Papers of Albert Einstein. Not incidentally, most of the editors of those volumes are also the main contributors here. Volume I in fact contains a facsimile reproduction and transcriptions of the notebook. Volume II features a line by line analysis by Michael Janssen, J\"urgen Renn, Tilman Sauer, John Norton, and John Stachel. Amazingly, they are able to follow definitively almost every step in Einstein's search for field equations. Where necessary they fill in missing calculations. Considered as an historical epistemological laboratory I would guess that resources offered here are without parallel in the history of science. Volume I contains a precise magisterial summary by Renn and Sauer of the metaprocesses that steered Einstein to his conclusion. Relativists will likely want to read the specifics in the second volume first, as I did, before tackling this more generalized cognitive overview. 

Einstein had already in 1912 learned an important lesson regarding gravitational field equations. The natural candidate for his uniform static field was Poisson's equation $\nabla^2 c = k c \rho$. (He required linearity in $c$ from the physical requirement that $c$ should only be fixed up to a constant factor. So in this theory Newton's gravitational constant is understood not to be a constant after all.) But, as discussed by Renn and Sauer, this equation is inconsistent with momentum conservation. Since this becomes one of Einstein's crucial physical checks through the period documented by the Notebook, I will give the argument. The force per unit volume in terms of the mass density $\rho$ is $f_a = k^{-1} \rho \partial_a c$. Substituting from the field equations for the source we find $f_a  = \frac{1}{kc} (\nabla^2 c) \partial_a c$. But this must be expressible as a total divergence, otherwise (supposing that $c$ falls off appropriately at infinity) the net force will not be zero. In a straightforward calculation he then found the term that had to be added to the field equations so that this substitution yielded a total divergence.  The resulting field equation is $\nabla^2 c = k c \rho + \frac{1}{2c} \partial_a \partial_a$. It is noteworthy that later on the analogues of this second term containing first derivatives of the metric will be recognized and sought by Einstein as contributions from a gravitational stress energy tensor.  In fact, Einstein's immediate task in Zurich was to formulate a field equation for the ten-component metric field, recognizing that material sources needed to be represented by a stress energy tensor. The mathematical problem he faced was to find a generalization of a Laplacian acting on a metric field. And this generalized expression had to admit sufficient covariance to at least include transformations to uniformly accelerated frames. He even wrote down already at this stage in 1912 the variation of the spacetime line element, in terms of a ten-component metric in arbitrary coordinates, yielding the free particle equation of motion. He already realized that the variation of the spacetime line element, in terms of a ten-component metric in arbitrary coordinates, yields the free particle equation of motion. This suggested to him that he could generalize to the equations of motion of a pressureless dust, derived from the same variational principle. This led, in part through inspired guesswork analyzed elsewhere by Norton, to the relation that would constitute the core of all subsequent calculations. He found an expression for the four-force per unit volume, or force density, experienced by matter described by a stress energy tensor $T^{\mu \nu}$.  Due to its appearance in the force term on the left hand side, and the total divergence on the right hand side, the equivalence principle is assured:
 \beq
 \frac{1}{2} \sqrt{-g} \frac{\partial g_{\mu \nu}}{\partial x^\beta} T^{\mu \nu}  =  \partial_\alpha \left(  \sqrt{-g} g_{\beta \mu}T^{\mu \alpha} \right)  \label{standard}
  \eeq
Einstein deduced that this equation transforms tensorially under arbitrary spacetime coordinate transformations provided $T^{\mu \nu}$ is symmetric. He interpreted (\ref{standard}) as an expression for energy momentum balance. The right hand side is the time rate of change of material energy momentum. It did not escape Einstein's notice that  (\ref{standard}) is the electromagnetic analogue of the Lorentz force equation, and indeed, continuing this analogy, the right hand side should be minus the time rate of change of the energy momentum density of the gravitational field. Carrying this electromagnetic conceptual framework further (corresponding to mental models identified by the authors), the task is then to find the field engendered by the material source.

He attempted to learn how to construct adequately covariant objects by studying the first and second order Beltrami operators acting on scalars. Incidently,  in the process he came remarkably close to developing an understanding of covariant derivatives that, as will probably surprise most relativists, was first enunciated later by Levi-Civita in 1917.  He found the expression for the derivative of a second rank symmetric tensor by transforming from a locally flat spacetime to an arbitrary coordinate system. A strategy he developed in this investigation became a standard calculational tool. He assumed that a given quantity transformed covariantly and then found self-consistency conditions that needed to be satisfied. Eventually he turned to the simpler task of exploring the covariance properties of these conditions rather than those of the original object.  Einstein finally turned to his friend and now colleague at the ETH, Marcel Grossmann, for assistance. Grossmann brought the Riemann-Christoffel tensor to his attention. The two initially attempted to find field equations using the Ricci tensor. They were well aware that this expression was generally covariant. But Einstein insisted that several physical requirements be fulfilled by any candidate field equation. It is the complicated, and almost always contradictory interplay between the mathematical emphasis on covariance, and the physical requirements of equivalence principle, energy momentum conservation, and the correspondence limit that highlight the ensuing drama.  Decisive at this stage was the belief that the metric of the static correspondence limit was spatially flat, and therefore in this limit the field equations should become simply $\eta^{\rho \sigma} \partial_\rho \partial_\sigma g_{\mu \nu} = \kappa T_{\mu \nu}$.  This was achieved through imposing the so-called harmonic condition/restriction $g^{\mu \nu} \Gamma^\alpha_{\mu \nu} = 0$. In addition energy momentum conservation was attained in the weak field limit by imposing the so-called Hertz condition/restriction $\partial_\nu g^{\mu \nu} = 0$.

Here a short interlude is in order. I am purposely circumspect on the designation of this relation, reflecting what I regard as a healthy continuing debate amongst the authors. Renn and Sauer introduced the notion of ``restriction'' in situations such as this to signify, from our modern perspective, that Einstein was intending to restrict the set of physical solutions of equations of greater covariance. And there are clear instances in the notebook  where this is the case. On the other hand, Norton proposes what, as far as I can tell, is a minority thesis that through the period leading up to the so-called Entwurf (outline) theory published by Einstein and Grossmann in 1913, they were consciously applying coordinate conditions. In other words, the Hertz condition was viewed by them as a choice of coordinate system, as is indeed any good modern gauge choice. But we view any remaining gauge freedom as a failure to uniquely fix the coordinates, meaning that there will be non-trivial equivalence classes of solutions that all correspond to the same spacetime.  According to Norton on the other hand, Einstein and Grossmann viewed the remaining freedom as representing relativity transformations between equivalent coordinate systems. There is much that is attractive about this proposal. It does seem to accord well with some of Einstein's otherwise enigmatic statements. For example, in 1914, in defense of the Entwurf theory that is covariant only under general linear transformations, Einstein  writes that there must indeed exist (in this case unknown) generally covariant equations   that become the Entwurf equations with the imposition of an appropriate coordinate condition. He insists that otherwise the equations would not relate physical quantities. We could interpret this in our modern sense as meaning they assist in identifying physically distinct solutions.  For reasons that I will outline below, Norton interprets Einstein's errant view of coordinate conditions as a consequence of his unwittingly according reality to spacetime coordinates.

But let us return to the main narrative. The two conditions encountered above led to the requirement that the trace of the stress energy must vanish. Although Einstein would toy with this idea again in 1915, briefly embracing the idea advanced by Gustav Mie that all interactions were fundamentally electromagnetic, at this time the vanishing trace was physically inadmissible. On the other hand the addition of a trace of the linearized metric on the left hand side of the field equations eliminated this condition - but thereby spoiling the static limit! Significantly, these rejected field equations were the exact linearized version of the general theory published in November, 1915. The next idea was to restrict to covariance under unimodular coordinate transformations (satisfying $\left| \frac{\partial x'}{\partial x} \right| = 1$). The determinate of the metric transforms as a scalar under these transformations, and it was possible to split off two terms from the Ricci tensor as a candidate left hand side of the field equations, resulting in
$ \partial_\rho \Gamma_{\mu \nu}^\rho - \Gamma_{\mu \sigma}^\rho \Gamma_{\rho \nu}^\sigma = \kappa T_{\mu \nu}. 
$
These field equations were also rejected, but in a series of steps analyzed by Janssen and Renn, Einstein slowly and steadfastly worked his way back to them and published them in the tumultuous month of November, 1915. Hence the authors call this the November theory. The evidence presented in their essay overturns older conventional scholarship, in particular that of Abraham Pais. It also conflicts with Einstein's later recollection, according to which he eventually simply gave up on his physical strategy and gave precedence to the mathematical requirement of covariance. According to Janssen and Renn, and in consonance with Einstein's own recollection of his ``fateful error'' in not recognizing that the Christoffel symbol was the true representative of the gravitational field, it was through sustained application of the physical strategy that Einstein eventually corrected this error. 

The foundations of this physical strategy were established while massaging and eventually dismissing the November theory, and during the creation and elaboration of the Entwurf theory. Einstein inserted candidate field equations into the force density equation, looking for corrections that would both identify a suitable gravitational stress energy tensor and result in the vanishing of the the divergence of the total stress energy.  Eventually after having failed beginning with the November tensor to find field equations that admitted the metric of a uniformly rotating Minkowski frame as a solution, even after having substantially simplified the physical conditions through the clear use of a broader coordinate restriction, Einstein abandoned the idea of beginning with expressions of broad covariance. The Entwurf theory gave the physical criteria highest priority. He simply started with a modest ``core'' field equation and developed a clever method that insured both energy momentum conservation and the ``correct'' static limit. All of this is documented in the Notebook. Norton argues that it was in the transition to the Entwurf theory that Einstein abandoned coordinate conditions. He believes that it was the failure of the November theory to include the rotation metric that induced this change. In his opinion, Einstein believed not only that the coordinate condition singled out a system of coordinates, but also that all solutions were conceived by Einstein as existing in the ``same'' coordinate system (analogous to inertial frames in special relativity). As corroborating evidence he cites Einstein's remarks in a 1916 letter to Ehrenfest in which he writes ``...I myself needed so long to achieve full clarity on this point. Your problem has its roots in that you instinctively treat the reference system as something real.''  If the implication is correct that Einstein had himself acted in this manner, then one can understand why the November theory was discarded.  Whereas the rotation metric is obtainable through a unimodular transformation from Minkowski space, it is not a solution in the one and only admissible coordinate system.  Indeed, Norton argues that this unconscious assignment of reality to the coordinate system predates and is consistent with a similar argument that Einstein forwarded in 1914 in rejecting generally covariant field equations. His so-called ``hole argument'' showed that the metric field at a given coordinate point is not uniquely fixed by the material stress energy tensor. Of course, he later withdrew this objection, recognizing that only material and/or field coincidences fix spacetime points. Coordinates have no independent reality.

According to Renn and Janssen, it was only after Einstein had developed a powerful variational technique while elaborating the Entwurf theory that he finally was in position to  select the correct gravitational field. This method established a rigid connection between energy momentum conservation and covariance. In fact it inspired Emmy Noether's 1918 theorems. Einstein's earlier modifications of the field equations constituted evidence that such a relation existed.  In elaborating the Entwurf theory in 1914 he found that the same four conditions conditions enforced both linear covariance and energy momentum conservation, but a sloppy error in his variational procedure initially obscured the relation between additional compulsory conditions. He found, in analogy with electromagnetism, that the Entwurf Lagrangian was quadratic in the quantity that he had taken as his gravitation field, $\frac{1}{2} g^{\beta \rho} g_{\rho \alpha, \mu} $. But he was not satisfied with the Entwurf theory.  The primary concern seems to have been that the Entwurf theory did not predict the correct perihelion advance for the planet Mercury (a problem that had already been pointed out to him two years earlier by his friend and collaborator Michele Besso - as discussed in an essay by Janssen). So at this stage he possessed the mathematical tool, born out of his endeavor to enforce his physical constraints, that enabled him to work out in a straightforward manner the consequence of replacing $\frac{1}{2} g^{\beta \rho} g_{\rho \alpha, \mu} $ by the Christoffel tensor.
It is remarkable that a documentation analogous to the Zurich notebook does exist for this period; the evolution of Einstein's reasoning was published in November, 1915, in a series of four papers submitted to the Prussian Academy of Science! Einstein did not initially apply the new variational technique in recovering the November theory. But he did use lessons he had learned from this formalism to disentangle the static limit constraint from the conditions necessary for covariance and energy momentum conservation. In so doing he arrived at our modern understanding of coordinate conditions.

Unfortunately, due to lack of space, I cannot adequately address the numerous additional sources and analyses that appear in these volumes. They establish the early twentieth century scientific context, examine Einstein's significant interaction with his physics colleagues, and track the further elaboration of the general theory and its mathematical spinoff in the period through 1919. Renn's analysis of Mach's influence on Einstein deserves special mention, as does the plausible mathematical historical fantasy by Stachel, Julian Barbour's insightful observations on Einstein's failure to incorporate a model of rods and clocks into his theory, and the illuminating discussions by Renn and Norton on Abraham and Nordstr\"om, respectively. For those who might still believe that Hilbert is the real hero of this story, I recommend the analyses of the Hilbert's first draft by Renn, Stachel, and Sauer. Finally, the editors are to be applauded for assembling together for the first time English translations of several of the key contributions by Einstein's contemporaries to the development of general relativity. Besides the work of Abraham, Nordstr\"om, and Mie, included among these gems are path-breaking articles by Lorentz, F\"oppl, Schwarzschild, Poincar\'e,  Born, Hilbert, Grassmann, Levi-Civita, Weyl and Cartan.

It will not have escaped the reader's attention that Einstein's insistence on energy momentum conservation is highly problematical from our modern perspective. And herein lies the mystery and the wonder of the epistemological process that has been brought to light by the authors. Einstein's heuristics were firmed fixed within the mental framework of early twentieth century theoretical physics. To be sure, there did exist three largely disjoint disciplines, namely mechanics, electromagnetism, and thermodynamics. And depending on one's heritage, the notions from one discipline might take precedence in attempting reconciliation in areas of overlap. Einstein was a largely self-taught master in all three, so he was perhaps more agile in his ability to shift and reinterpret emphases in response to experimental and theoretical challenges. We see clearly that he acted on the basis of a world view consonant with physics as understood at the time. Idealized particles from mechanics interacted with fields that were still not yet entirely liberated from their origins as alterations of the ether. Fields in the Lorentz model were determined by particle sources. Particle motion could itself be understood as constrained motion not entirely inconsistent with the natural motions postulated by Aristotle.  Energy conservation featured in every discipline, and it found its most elegant mathematical expression in the laws of thermodynamics. This view of nature constituted a mental frame for Einstein, and his operations within this framework have been exhaustively documented by the authors. It is only within this structure that one can make sense of the transformation that Einstein achieved, one might almost say against his will. Or perhaps more precisely stated, we are to understand this process as the realization of an ineluctable law of knowledge development.  In this particular process the equivalence principle reversed its position from its status as a curious secondary property of Newton's theory to a foundational principle. Along the way its meaning mutated to the extent that gravitational mass no longer exists. Einstein intended to extend the principle of relativity beyond the Poincar\'e transformations of inertial frames of reference, frames in which the coordinates enjoy a direct relation with measurable physical quantities. But in the final theory spacetime coordinates have lost their ontological significance and today the debate continues whether the ultimate theory is really a theory of relativity at all. Perhaps most remarkably, Einstein's unflagging insistence that energy momentum must be conserved led to repeated new insights and reformulations, frequently even in response to remarks of friends and critics, and the final result was a theory in which, as already recognized explicitly in print by Einstein in 1918, no meaning can be assigned locally to the gravitational stress energy tensor! 

I would venture that the lesson for us all is that we master to the best of our ability the physics handed to us by our forebears. Then having caught the glimmer of a unifying idea, we cling tightly, but always open to revision and reordering as events warrant. I strongly urge that you seek stimulation in this search from this masterful study of the genesis of general relativity.

\medskip

\noindent Donald Salisbury\footnote{Permanent address: Department of Physics, Austin College, Sherman, Texas 75090, USA. E-mail: dsalisbury@austincollege.edu}

\noindent Max Planck Institute for the History of Science

\noindent Boltzmannstra§e 22

\noindent 14195 Berlin, Germany

\noindent E-mail: dsalisbury@mpiwg-berlin.mpg.de

\end{document}